\documentclass[aps,prb,amsmath,amssymb,reprint,floatfix,superscriptaddress,longbibliography]{revtex4-2}
\usepackage{graphicx}
\usepackage{dcolumn}
\usepackage{wasysym}
\usepackage{bm}
\usepackage{color}
\begin{document}

\title{Inferring the dynamics of ionic currents from recursive piecewise data assimilation of approximate neuron models}

\author{Stephen A. Wells}
 \affiliation{Department of Physics, University of Bath, Bath BA2 7AY, United Kingdom}

\author{Joseph D. Taylor}
 \affiliation{Department of Physics, University of Bath, Bath BA2 7AY, United Kingdom}

\author{Paul G. Morris}
 \thanks{Present address: Department of Physiology, Development and Neuroscience, University of Cambridge, Cambridge CB2 3DY, United Kingdom}
 \affiliation{Department of Physics, University of Bath, Bath BA2 7AY, United Kingdom}

\author{Alain Nogaret}
 \email{A.R.Nogaret@bath.ac.uk}
 \affiliation{Department of Physics, University of Bath, Bath BA2 7AY, United Kingdom}

\begin{abstract}
We construct neuron models from data by transferring information from an observed time series to the state variables and parameters of Hodgkin-Huxley models.  When the learning period completes, the model will predict additional observations and its parameters uniquely characterise the complement of ion channels.  However, the assimilation of biological data, as opposed to model data, is complicated by the lack of knowledge of the true neuron equations.  Reliance on guessed conductance models is plagued with multi-valued parameter solutions.  Here, we report on the distributions of parameters and currents predicted with intentionally erroneous models, over-specified models, and an approximate model fitting hippocampal neuron data.  We introduce a recursive piecewise data assimilation (RPDA) algorithm that converges with near-perfect reliability when the model is known.  When the model is unknown, we show model error introduces correlations between certain parameters.  The ionic currents reconstructed from these parameters are excellent predictors of true currents and carry a higher degree of confidence, $>95.5\%$, than underlying parameters, $>53\%$.  Unexpressed ionic currents are correctly filtered out even in the presence of mild model error.  When the model is unknown, the covariance eigenvalues of parameter estimates are found to be a good gauge of model error.  Our results suggest that biological information may be retrieved from data by focussing on current estimates rather than parameters.
\end{abstract}

\maketitle

\section{Introduction}

Data assimilation is a first-in-class method for building models of nonlinear dynamical systems~\cite{Wachter2006,Boyd2004,Bergmann2016,hogg2011hsl,Petzold1983,Kuhn1951}.  It estimates parameters by synchronizing the model state variables to time series observations.  Once the training period is complete, the optimized models successfully predict complex time series oscillations ranging from chaotic dynamics~\cite{Kalnay2003,Parlitz2014b} to biological neuron~\cite{Nogaret2016,Meliza2014,Nogaret2022,Abarbanel2013} and networks~\cite{Armstrong2020}.  The estimated parameters hold the further promise of revealing hidden internal properties of biological systems.  The challenge, however, is that the state equations of neurons are generally unknown and modelling neurons with empirical models such as the Hodgkin-Huxley model~\cite{Lillacci2010,Pospischil2008,Bardsley2014,Achard2006,
Baldi1998,Meliza2014,Brookings2014,Nogaret2016} introduces model error which biasses parameter estimates away from true values~\cite{Pape1990,Gerstner2002,Moraes2013,Chan2010,Migliore2005,
Molyneaux2007,Zeng2017,Bina2010,Lerche2012}.  Although much attention has been drawn to multi-valued solutions in biological inference problems~\cite{Gutenkunst2007,Oleary2015,Goaillard2009,Prinz2004}, a systematic study is needed to understand how model error perturbs the distribution of parameters, and how it can be mitigated.  Correcting parameter sloppiness is complicated by the additional requirements that models have to meet such as observability~\cite{Takens1981,Schumann2011,Schumann2016,Aeyels1981} (the state vector is uniquely constrained at every point of the assimilation window) and identifiability~\cite{Csercsik2012,Raman2017,Raue2010,Parlitz2014a, Parlitz2014b,Letellier2006,Letellier2005} (the stimulation protocol elicits enough information from the neuron to constrain all model parameters).  These conditions can be difficult to quantify and are not systematically checked~\cite{Getting1989,Jing1999,Rubin2009,Marder2001,Nogaret2016,Selverston2010}. As a result, several authors have opted to develop leaner models with fewer parameters ~\cite{Clark2022,Otopalik2017,Hillel2018} in an attempt to reduce parameter error, or simply focused on predictions alone~\cite{Gutenkunst2007}.

Here, we show that ionic current predictions provide more reliable information on the biology than parameters and we propose that the covariance matrix of parameter estimates constitutes a suitable metric of model error when the biological model is unknown.  We introduce Recursive Piecewise Data Assimilation (RPDA) as a novel parameter search algorithm.  RPDA biases the state vector towards the true solution by re-injecting membrane voltage data in the state vector at regular intervals across the assimilation window.  When the model is known, RPDA obtains the true solution with near perfect reliability and minimal error ($<0.1\%$), independently of the choice of initial state vectors and stimulation protocols.  The uniqueness of solutions in turn establishes that observability and identifiability criteria are fulfilled.  We then introduce model error by detuning a gate exponent and study the dispersion of parameters estimated by two variants ($ErrM1$, $ErrM2$) of the model used to generate the data.  Two further model variants adding a supplementary ion channel to the original model ($ErrM3$) and to its $ErrM1$ variant ($ErrM4$) gave the parameter and current dispersions of over-specified models.  We find that model error introduces correlations between blocks of parameters defining the same ion channel type.  The calculation of ionic currents integrates these correlations with the effect that mean current predictions deviate by less than 8\% from their true values compared to over 100\% for some parameters.   Uncertainty on current estimates is less than 4.5\% whereas it is 47\% on parameters.  Over-specifying conductance models by adding extra ion channels to them is not found to induce significant compensation between parameters.  Unexpressed ion channels are correctly filtered out while those present are assigned the correct current values, even by the mildly erroneous model $ErrM4$.  We finally compare the uncertainty on currents estimated with intentionally erroneous models ($ErrM1$-$ErrM4$) to those of a guessed CA1 model trained on CA1 hippocampal neuron data.  We find that the uncertainty on the reconstructed CA1 currents is 14\% compared to 4.5\% on $ErrM2$ currents.  As the uncertainty on ionic currents increases with model error, we estimate that the error in the CA1 model is about 3 times greater than the error introduced by detuning a single gate exponent in $ErrM2$.  Our findings suggest that the predictive accuracy of RPDA is sufficient to detect ion channel dysfunction~\cite{Kagan2002,Savio2011,Duda2016,Brown2011,Shreaya2014,Nurse2010,Dreyer2013,Oh2012,Nadim2014} and to infer the action of ion channel antagonists~\cite{Morris2023}.

\section{Recursive Piecewise Data Assimilation}

We use data assimilation~\cite{Boyd2004,Wachter2006} to optimize the state variables and parameters of Hodgkin-Huxley type models~\cite{HH1952} by minimizing a least-square cost function~\cite{Nogaret2016,Brookings2014,Meliza2014}.  This function measures the misfit between a state variable representing the membrane voltage, $x_1$, and the experimental membrane voltage, $V_{mem}$, recorded at discrete times $t_i$, $i\in[0,N]$ spanning the assimilation window of duration $T$:

\begin{equation}
c(\vec{x}(0)) = \frac{1}{2}\sum_{i=0}^{N} \left[ (x_1(t_i,\vec{x}(0))-V_{mem}(t_i))^2 + x_{L+1}(t_i)^2    \right].
\label{eq:eq1}
\end{equation}

\noindent The state of the neuron is described by a vector $\vec{x}$ with $L+K+2$ components.  Vector components are $x_1$, the membrane voltage; $x_2,\dots,x_L$, the gate variables of ion channels; $x_{L+1}$ and $x_{L+2}$, the Tikhonov regularization variable~\cite{Tikhonov1943,Abarbanel2010,Creveling2008} and its time derivative; and $x_{L+2}, \dots, x_{L+K+2}$, the model parameters.  The model parameters do not depend on time.  The cost function is minimized subject to both equality constraints:

\begin{equation}
\frac{dx_l}{dt}=F_l(x_1,\dots,x_{L+K+2}) \;\;\; {\scriptstyle l=1,\dots,L+K+2},
\label{eq:eq2}
\end{equation}

\noindent specified by the neuron model and the zero time derivatives of model parameters expressed as follows:

\begin{equation}
F_l(\vec{x}) = \left\{
\begin{array}{cl}
-J/C -x_{L+1}\left(x_1-V_{mem}\right), & {\scriptstyle l=1} \\
\left[x_{l\infty}-x_l\right]/\tau_l, & {\scriptstyle l=2,\dots,L} \\
x_{L+2}, & {\scriptstyle l=L+1} \\
\mbox{unspecified}, & {\scriptstyle l=L+2} \\
0, & {\scriptstyle l=L+3,\dots,L+K+2}
\end{array}
\right.
\label{eq:eq3}
\end{equation}

\noindent and inequality constraints:

\begin{equation}
x_l^{min} \leq x_l \leq x_l^{max} \;\;\; {\scriptstyle l=1,\dots,L+K+2},
\label{eq:eq4}
\end{equation}

\noindent specifying the range of variation of membrane voltage, gate variables, regularization term and parameters,

\begin{equation}
\left[x_l^{min},x_l^{max}\right] = \left\{ \begin{array}{cccl}
-100mV & , & 50mV & {\scriptstyle l=1} \\
0 & , & 1 & {\scriptstyle l=2,\dots, L} \\
0 & , & 1 & {\scriptstyle l=L+1} \\
-1 & , & 1 & {\scriptstyle l=L+2} \\
p_l^{min} & , & p_l^{max} & {\scriptstyle l=L+3,\dots,L+K+2} . \end{array}
\right.
\label{eq:eq5}
\end{equation}

\noindent $J=J_{Na}+J_K+\dots-J_{inj}$ is the current per unit area of the neuron membrane.  This includes all ionic currents (Na, K ...) and the current injected to drive neuron oscillations, $J_{inj}$.  $C$ is the membrane capacitance, $\tau_l$ the recovery time of ionic gate $l$, and $x_{l\infty}$ is the steady-state value of gate variable $x_l$.  The user sets the parameter search range, $[p_l^{min},p_l^{max}]$, to the widest biologically plausible range for each parameter.  Data assimilation outputs the optimal parameters and the state variables at $t=0$ as $\vec{x}(0)$.

The convergence of data assimilation is compromised by badly conditioned problems.  When the model is known, convergence may fail when the observability and identifiability criteria are unmet or, if the conductance model is large ($L>12$), when the parameter search risks getting stuck in a local minimum of the cost function.  These issues have been partially remedied by either increasing the embedding space~\cite{Aeyels1981,Schumann2011,Schumann2016,Parlitz2014b,Letellier2005,Rey2014}, improving the design of the injected current waveform, or using noise regularization~\cite{Bardsley2014,Taylor2020}.  In addition, when the model is unknown, as in real neurons, the global minimum may not coincide with the true solution.  Parameter searches starting from different initial state vectors may get stuck at different minima, giving multi-valued solutions.

\begin{figure*}
\includegraphics[width=\linewidth]{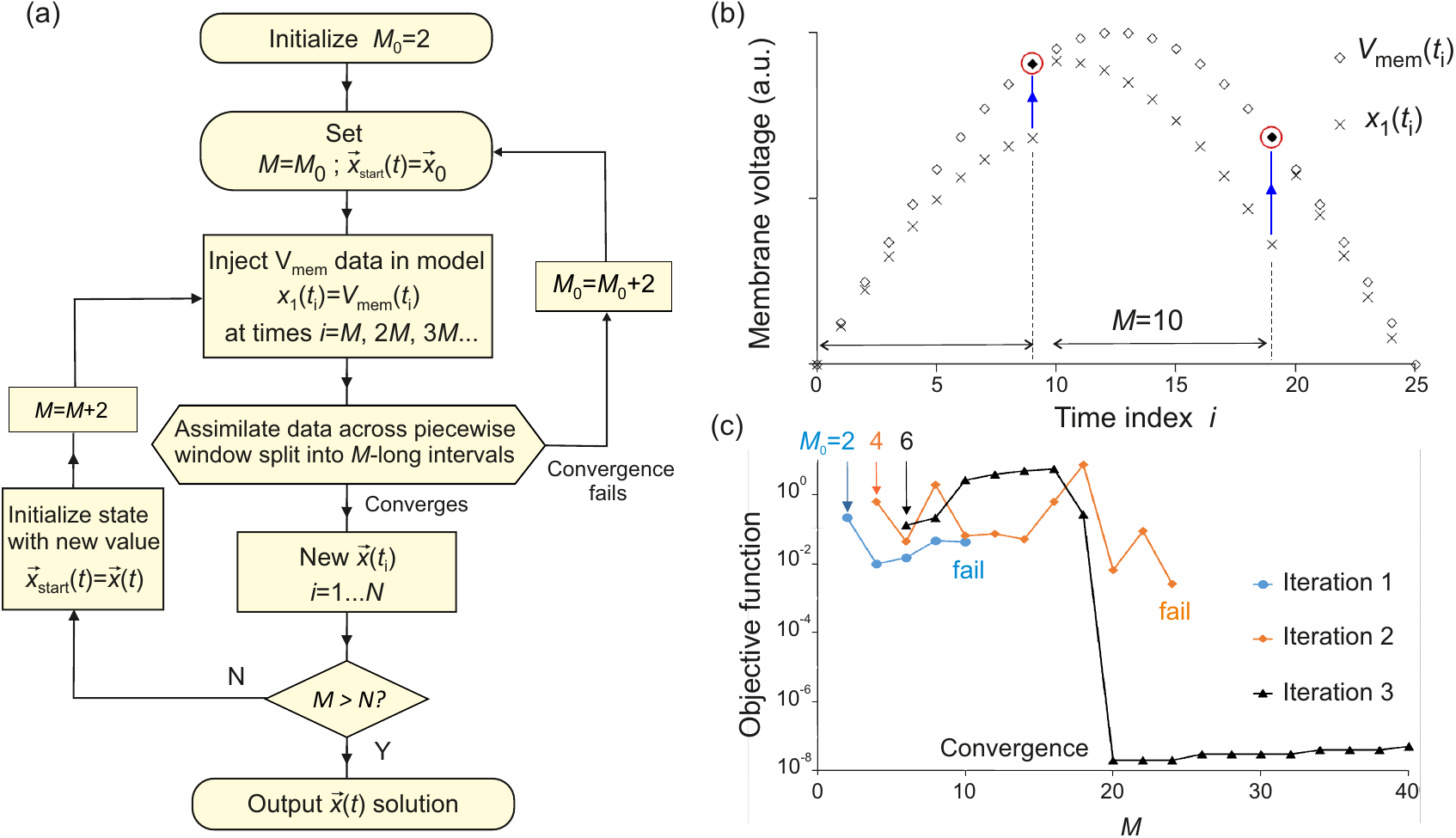}
\caption{\textbf{RPDA algorithm}
\\
(a) Algorithm flowchart; (b) The RPDA algorithm synchronizes the membrane voltage variable $x_1$ to the time series data, $V_{mem}$, in blocks of $M$ discrete time points.  The algorithm re-injects $V_{mem}$ in the model at the beginning of each block (red circle); (c) Example of recursive convergence starting from $M_0=6$ and increasing the block size to $M=N$.  This follows two ``false starts'' at $M_0=2$ and $M_0=4$.}
\label{fig:fig1}
\end{figure*}

The Recursive Piecewise Data Assimilation (RPDA) algorithm, we present in Fig.\ref{fig:fig1}a, resolves these issues by biassing the parameter search towards the local minimum closest to the true solution in the ill-posed case, and towards the global minimum in the well-posed case.  The parameter search is biased by re-injecting membrane voltage data into the model every $M$ data-points.  This is done by substituting $x_1$ with $V_{mem}$ at time points $t_0, t_{M-1}, t_{2M-1} \dots $ in the linearized expression of the equality constraints~\cite{Taylor2020}:

\scriptsize
\begin{eqnarray}
\begin{aligned}
&x_l(t_{i+2}) = x_l(t_{i})+\Delta t \left[\frac{1}{3}F_l(\vec{x}(t_i))+\frac{4}{3}F_l(\vec{x}(t_{i+1}))+\frac{1}{3}F_l(\vec{x}(t_{i+2})) \right], \\
&x_l(t_{i+1}) = \frac{1}{2}\left(x_l(t_i)+x_l(t_{i+2})\right)+\frac{\Delta t}{4}\left[F_l(\vec{x}(t_i))-F(\vec{x}(t_{i+2}))\right], \\
& \;\;\;\;\;\; {\scriptstyle l = 1,\dots,L+K+2,\;\;\;\;\;\; i=0,\dots,N-2}.
\end{aligned}
\label{eq:eq6}
\end{eqnarray}
\normalsize

\noindent where $\Delta t = T/N$.  At other times the state vector is propagated normally from $t_i$ to $t_{i+2}$ by Eqs.\ref{eq:eq6}.  The substitution of $V_{mem}$ also impacts the first and second derivatives of the objective function with respect to $x_1$.   Where the cost function ceases to depend on $x_1$, its derivatives with respect to $x_1$ vanish.  We explicitly set these derivatives to zero at times $t_0, t_{M-1}, t_{2M-1} \dots$.  This produces a discontinuity in $x_1(t)$ at the beginning of each $M$-block (Fig.\ref{fig:fig1}b) which is the tradeoff for imposing the bias to the solution.  The RPDA method has similarities with the multiple shooting approach proposed by Zimmer and Sahle~\cite{Bergmann2016,Zimmer2015}, which also fits data in a piecewise manner.  However our RPDA method performs piecewise assimilation of blocks of data through \textit{redefined constraints}, whereas the Zimmer and Sahle's approach \textit{redefines the cost function}.

The RPDA algorithm recursively improves the accuracy on parameter estimates by relaxing the bias and re-injecting data over longer $M$-intervals, while restarting the parameter search from the previous estimate.  The initial block size is $M_0=2$.  The estimated state vector is then used as the new initial state in the next parameter search when $M$ is incremented from $M_0$ to $M_0+2$.  The search terminates when $M$ becomes greater than $N$ ($N=50,001$).  The iterations ultimately restore the continuity of $x_1(t)$ across the assimilation window.  When the initial bias is too strong the parameter search may fail.  The algorithm then restarts the parameter search from the next larger block size ($M_0=4$).  This process is depicted in Fig.\ref{fig:fig1}c where the parameter search is restarted twice after convergence failed starting from $M_0=2$ then $4$ to succeed with $M_0=6$.

Previous work~\cite{Bardsley2014,Taylor2020} has explored the use of additive noise as a regularization method, showing that the stochastic perturbation of the fitting landscape by noise can disrupt local minima and improve convergence to the global minimum. In the RPDA method, each segment of length $M$ similarly perturbs the fitting landscape.  RPDA has the further advantage that the solution will remain a good solution for any value of $M$ thanks to the bias from re-injected data.

\section{Results}

We then examined the accuracy of RPDA in three cases of model error in the assimilating model: (i) the model is known and is correct; (ii) the model is erroneous but error is known; (iii) the model is erroneous and the error is unknown.  In the first case RPDA converges 100\% of the time towards to the true solution.  In the second case, we assimilate the same data with 4 variants of the original model incorporating either an erroneous gate exponent, redundant ion channels, or a combination of both.  In the third case, we use a guessed conductance model to assimilate membrane voltage recordings of hippocampal neurons.

\begin{figure*}
\includegraphics[width=\linewidth]{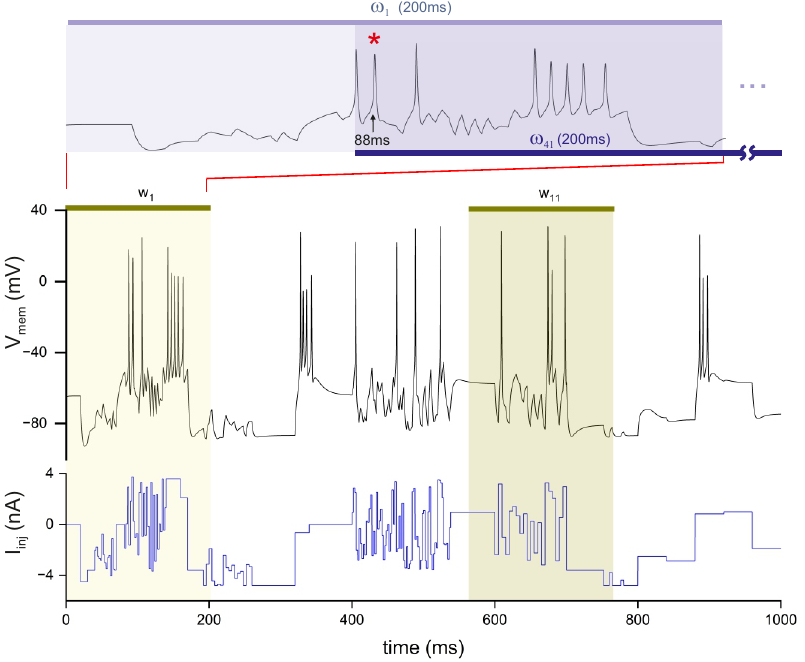}
\caption{\textbf{Neuron oscillations and assimilation ranges}
\\
Membrane voltage (black trace) generated by forward integration of the injected current protocol (blue trace) with the original RVLM model.  $w_1, \dots, w_{11}$ are the 11 assimilation windows used to generate the data in Table~\ref{tab:tab1} from well-posed problems.  Each window is 200ms wide (yellow bands) and offset from the next by 40ms.  \textit{Top panel}: $\omega_1, \dots, \omega_{41}$ are the 41 assimilation windows used the generate the data in Table~\ref{tab:tab2} from ill-posed problems.  These windows are also 200ms long but offset by 2ms so that all encompass the same action potential labelled (*) at 88ms.}
\label{fig:fig2}
\end{figure*}

\subsection{Well–posed problems}

Our reference model is a single compartment model of the rostral ventrolateral medulla (RVLM).  This model is exemplar of the models used to predict neuron oscillations~\cite{Meliza2014,Nogaret2016}.  It comprises five common types of ion channels: transient sodium (NaT), delayed-rectifier potassium (K), low threshold calcium (CaT), hyperpolarized cyclic nucleotide (HCN), and leak (Appendix A).  The model has $L=7$ state variables and $K=40$ adjustable parameters whose reference values, $p_k$ ($k=l-(L+2)$), are listed in Table.\ref{tab:tab1}.  The RVLM model configured with these parameters was used to forward-integrate the current protocol $I_{inj}(t)$ in Fig.\ref{fig:fig2} and generate voltage oscillations $V_{mem}(t)$ over 1000ms.  The sequence of depolarising and hyperpolarising steps of varying amplitudes and durations was designed to elicit a response from all ionic currents in the neuron.

\subsubsection{Uniqueness of solutions}

The convergence of RPDA was probed by initializing the state vector $\vec{x}$ at 28 different locations: 18 initial parameter values were chosen at random in the parameter search range, the other 10 had parameters set every 1/10th of the search interval.  In this way, the initial state vectors mapped the entire parameter space allowing the parameter search to approach the solution from different directions.  All assimilation runs used the same 200ms long data set labelled $w_1$ in Fig.\ref{fig:fig2}.

RPDA successfully converged to the true solution from all 28 initial conditions (100\% convergence rate).  23 of these assimilation runs converged from $M_0=2$. 3 more required one restart from $M_0=4$.  The last 2 required two consecutive restarts from $M_0=4$ and $6$, as shown in Fig.\ref{fig:fig1}c.  The mean block size at which parameter search arrived within 0.1\% of true values was $\langle M \rangle=22$.  This shows convergence is rapid once the algorithm has selected a suitable starting block size $M_0$.

The mean values of parameter estimates, $\mu_k$, and their standard deviations, $\sigma_k$, were calculated over the 28 sets (Table.\ref{tab:tab1}).  A majority of $\mu_k$ (34/40) are within $<0.1\%$ of the true parameter values.  Outliers such as $t_3$, $t_5$, $t_7$, $\epsilon_7$ relate to gate recovery times.  They nevertheless remain within $1\%$ of the true values.  In general the ultra narrow dispersions ($\sigma_k/\mu_k < 0.01\%$) confirm that parameters are fully constrained and the system is observable.

\begin{table*}[t]
\begin{center}
\begin{tabular}{|c c c|c|c c c|c c c|} % <-- Alignments: 1st column left, 2nd middle and 3rd right, with vertical lines in between
\cline{4-10}
\multicolumn{3}{c|}{ } & \multicolumn{1}{c|}{True} & \multicolumn{3}{c|}{28 initial guesses} & \multicolumn{3}{c|}{11 assimilation windows} \\
\hline
$k$ & \multicolumn{2}{c|}{Parameter}  & \;\;\;\;\; $p_k$ \;\;\;\;\; & \;\;\;\;\; $\mu_k$ \;\;\;\;\; & \;\; $\left|\frac{\mu_k-p_k}{p_k}\right|$ \;\; & \;\; $\left|\frac{\sigma_k}{\mu_k}\right|$ \;\; & \;\;\;\;\; $\mu'_k$ \;\;\;\;\; & \;\; $\left|\frac{\mu'_k-p_k}{p_k}\right|$ \;\; & \;\; $\left|\frac{\sigma'_k}{\mu'_k}\right|$ \;\;\\
\multicolumn{3}{|c|}{} & & & \scriptsize \% & \scriptsize \% & & \scriptsize \% & \scriptsize \% \\
\hline \hline
1. & $A$ & $\times0.1 $mm$^2$ & 0.290 & 0.290 & 0.001 & 0.0001 &  0.290 & 0.0030 & 0.0050 \\ \hline \hline
2. & $g_L$ & mS.cm$^{-2}$ & 0.465 & 0.465 & 0.002 & 0.0008 & 0.465 & 0.0073 & 0.0143 \\
3. & $E_L$ & mV  &  -65.00 & -64.99 & 0.0009 & 0.0001 & -65.00 & 0.0011 & 0.0045 \\  \hline \hline
4. & $g_{Na}$ & mS.cm$^{-2}$  &  69.00 & 68.91 & 0.1310 & 0.0013 & 69.16 & 0.2254 & 0.8875 \\
5. & $E_{Na}$ & mV & 41.00 & 41.01 & 0.0135 & 0.0001 & 40.99 & 0.0069 & 0.0200 \\  \hline
6. & $V_{t2}$ & mV & -39.92 & -39.92 & 0.0025 & 0.0002 & -39.92 & 0.0017 & 0.0082 \\
7. & $\delta V_2$ & mV & 10.00 & 10.00 & 0.0039 & 0.0002 & 9.99 & 0.0072 & 0.0293 \\
8. & $\delta V_{\tau 2}$ & mV & 23.39 & 23.38 & 0.0286 & 0.0009 & 23.39 & 0.0187 & 0.0260 \\
9. & $t_2$ & ms  &  0.143 & 0.143 & 0.1370 & 0.0023 & 0.143 & 0.0731 & 0.1797 \\
10. & $\epsilon_2$ & ms & 1.099 & 1.099 & 0.0066 & 0.0003 & 1.099 & 0.0156 & 0.0382 \\ \hline
11. & $V_{t3}$ & mV & -65.37 & -65.36 & 0.0095 & 0.0003 & -65.40 & 0.0411 & 0.1285 \\
11. & $\delta V_3$ & mV & -17.65 & -17.65 & 0.0124 & 0.0009 & -17.65 & 0.0187 & 0.0461 \\
13. & $\delta V_{\tau 3}$, & mV & 27.22 & 27.21 & 0.0148 & 0.0011 & 27.23 & 0.0202 & 0.0807 \\
14. & $t_3$ & ms & 0.701 & 0.701 & 0.0624 & 0.0007 & 0.701 & 0.0072 & 0.0662 \\
15. & $\epsilon_3$ & ms & 12.90 & 12.90 & 0.0174 & 0.0002 & 12.91 & 0.0906 & 0.2980 \\  \hline \hline
16. & $g_K$ & mS.cm$^{-2}$ & 6.90 & 6.91 & 0.0878 & 0.0080 & 6.90 & 0.0293 & 0.1265 \\
17. & $E_K$ & mV & -100.00 & -100.00 & 0.0029 & 0.0003 & -100.01 & 0.0066 & 0.0254 \\  \hline
18. & $V_{t4}$ & mV & -34.58 & -34.57 & 0.0140 & 0.0012 & -34.58 & 0.0127 & 0.0106 \\
19. & $\delta V_4$ & mV & 22.17 & 22.18 & 0.0462 & 0.0027 & 22.18 & 0.0267 & 0.0210 \\
20. & $\delta V_{\tau 4}$ & mV & 23.58 & 23.59 & 0.0241 & 0.0011 & 23.57 & 0.0307 & 0.0560 \\
21. & $t_4$ & ms & 1.291 & 1.292 & 0.0491 & 0.0051 & 1.291 & 0.0343 & 0.0650 \\
22. & $\epsilon_4$ & ms & 4.314 & 4.312 & 0.0524 & 0.0036 & 4.315 & 0.0269 & 0.1059 \\  \hline \hline
23. & $g_H$ & mS.cm$^{-2}$ & 0.150 & 0.150 & 0.0348 & 0.0051 &  0.150 & 0.0455 & 0.1587 \\
24. & $E_H$ & mV & -43.00 & -42.96 & 0.0861 & 0.0041 & -42.99 & 0.0329 & 0.0791 \\  \hline
25. & $V_{t5}$ & mV & -76.00 & -76.00 & 0.0006 & 0.0002 & -76.02 & 0.0220 & 0.0242 \\
26. & $\delta V_5$ & mV & -5.500 & -5.517 & 0.3099 & 0.0062 & -5.52 & 0.3440 & 0.5542 \\
27. & $\delta V_{\tau 5}$ & mV & 20.27 & 20.28 & 0.0338 & 0.0016 & 20.26 & 0.0306 & \textbf{1.960} \\
28. & $t_5$ & ms & 6.310 & 6.336 & 0.4088 & 0.0169 & 6.59 & 4.506 & \textbf{17.20} \\
29. & $\epsilon_5$ & ms & 55.05 & 55.03 & 0.0384 & 0.0013 & 54.67 & 0.6879 & \textbf{1.877} \\  \hline \hline
30. & $\bar{p}$ & $\mu$m.s$^{-1}$ & 0.1034 & 0.1030 & 0.3674 & 0.4924 & 0.1030 & 0.7145 & \textbf{5.115} \\  \hline
31. & $V_{t6}$ & mV & -65.50 & -65.49 & 0.0120 & 0.0006 & -65.49 & 0.0109 & 0.0154 \\
32. & $\delta V_6$ & mV & 12.40 & 12.39 & 0.0798 & 0.0032 & 12.41 & 0.0686 & 0.2568 \\
33. & $\delta V_{\tau 6}$ & mV & 27.00 & 27.12 & 0.4613 & 0.0170 & 27.04 & 0.1503 & 0.3330 \\
34. & $t_6$ & ms & 0.719 & 0.738 & 0.5071 & 0.1098 & 0.735 & 2.262 & \textbf{8.131} \\
35. & $\epsilon_6$ & ms & 13.05 & 13.06 & 0.0834 & 0.0072 & 13.03 & 0.1554 & 0.5307 \\ \hline
36. & $V_{t7}$ & mV  &  -86.00 & -86.00 & 0.0033 & 0.0298 & -85.98 & 0.0269 & 0.2461\\
37. & $\delta V_7$ & mV & -8.060 & -8.064 & 0.0436 & 0.0378 & -8.069 & 0.1164 & 0.2184 \\
38. & $\delta V_{\tau 7}$ & mV & 16.71 & 16.76 & 0.3021 & 0.0041 & 16.67 & 0.2465 & \textbf{1.311} \\
39. & $t_7$ & ms & 28.17 & 28.12 & 0.1802 & 0.0014 & 28.21 & 0.1389 & \textbf{1.366} \\
40. & $\epsilon_7$ & ms & 288.7 & 286.8 & 0.6526 & 0.2672 & 287.0 & 0.5688 & \textbf{2.347}  \\
\hline

\end{tabular}
\end{center}
\caption{\textbf{Dispersion of parameters estimated from different initial guesses and different assimilation windows} \\
The $p_k$ are the true parameters used to construct the data being assimilated (Fig.\ref{fig:fig2}).  The $\mu_k$ and $\sigma_k$ are the mean values and standard deviations of parameters estimated from different initial guesses of the state vector and assimilation window $w_1$.  The $\mu'_k$ and $\sigma'_k$ are the mean values and standard deviations of parameters estimated from assimilation widows $w_1, \dots, w_{11}$ starting from the same state vector.  Deviations from mean greater than 1\% are highlighted in bold.  We set $k=l-L-2$.}
\label{tab:tab1}
\end{table*}

With a $100\%$ convergence rate, the RPDA method improves over DA and DA with noise regularization which respectively achieve 67\% and 94\% convergence rates~\cite{Taylor2020}.  We next demonstrate that the current protocol fulfills the identifiability criterion by sampling different data sets.

\subsubsection{Identifiability}

Identifiability was investigated by probing the dispersion of parameters estimated from 11 different sections of the stimulation protocol.  These are the time windows labelled $w_1,\dots,w_{11}$ in Fig.\ref{fig:fig2}.  Each window is 200ms long (10,001 data points) starting at 0ms, 40ms, 80ms, $\dots$, 400ms.  The vector components of the initial state vector were set at the midpoint of the search range in all assimilation runs.

RPDA converged in all 11 assimilation windows.  The mean values of parameter estimates, $\mu'_k$, and standard deviations, $\sigma'_k$, are shown in Table~\ref{tab:tab1}.   Most parameters are very well constrained with 31/40 showing deviations of $\mu'_k$ from true values of less than 0.1\%.  Unsurprisingly, parameters related to gate kinetics show the largest standard deviations.  Yet the deviations of their mean values remain small suggesting parameters retain good predictive power.  For example the standard deviations on $t_6$ and $t_5$ are 8.1\% and 17\% respectively whereas the deviations of mean predicted values from true values are 2.2\% and 4.5\%.  Closer examination of the 11 parameter sets shows that the greater deviations occur in time windows which happen to include fewer action potentials, such as the 180-300ms interval in Fig.\ref{fig:fig2}.  Excluding these windows, the standard deviations fall in the normal range reported in the previous section.  This is a reminder of the importance of including a minimum of 5-6 action potentials in the assimilation widow for the identifiability condition to be fulfilled.

\begin{table*}[t]
\begin{center}
\begin{tabular}{|c c c|c c|c c|c c|} % <-- Alignments: 1st column left, 2nd middle and 3rd right, with vertical lines in between
\cline{4-9}
\multicolumn{3}{c|}{ } & \multicolumn{2}{c|}{Well-posed} & \multicolumn{4}{c|}{Ill-posed} \\
\cline{4-9}
\multicolumn{3}{c|}{ } & \multicolumn{2}{c|}{RVLM} & \multicolumn{2}{c|}{ErrM1} & \multicolumn{2}{c|}{ErrM2} \\
\hline
k & \multicolumn{2}{c|}{Parameter} & \;\;\;\; $\mu_k$ \;\;\;\; & \;\; $\left|\frac{\sigma_k}{\mu_k}\right|$ \;\; & \;\;\;\; $\mu_{1k}$ \;\;\;\; & \;\; $\left|\frac{\sigma_{1k}}{\mu_{1k}}\right|$\;\; & \;\;\;\; $\mu_{2k}$ \;\;\;\; & \;\; $\left|\frac{\sigma_{2k}}{\mu_{2k}}\right|$ \;\; \\
\multicolumn{3}{|c|}{} & & \scriptsize \% & & \scriptsize \% & & \scriptsize \% \\
\hline \hline
1. & $A$ & $\times 0.1$ mm$^2$ & 0.290 & 0.0038 & 0.291 & 0.1325 & 0.293 & 0.2814 \\ \hline \hline
2. & $g_L$ & mS.cm$^{-2}$ & 0.465 & 0.0107 & 0.4679 & 0.1104 & 0.4712 & 0.2812 \\
3. & $E_L$ & mV & -65.00 & 0.0353 & -65.26 & 0.3737 & -65.83 & \textbf{1.213} \\ \hline \hline
4. & $g_{Na}$ & mS.cm$^{-2}$ & 69.01 & 0.3138 & 40.68 & \textbf{2.794} & 19.21 & \textbf{1.260} \\
5. & $E_{Na}$ & mV & 41.00 & 0.0155 & 46.07 & \textbf{1.535} & 60.00	& 0.00 \\ \hline
6. & $V_{t2}$ & mV & -39.92 & 0.0033 & -37.81 & 0.1669 & -35.20 & 0.3431 \\
7. & $\delta V_2$ & mV & 9.998 & 0.0021 & 8.090 & 0.0476 & 5.256 & 0.0778 \\
8. & $\delta V_{\tau 2}$ & mV & 23.38 & 0.0041 & 20.76 & 0.1424 & 15.99 & 0.2437 \\
9. & $t_2$ & ms & 0.1429 & 0.0203 & 0.0920 & \textbf{1.681} & 0.0257 & 0.0425 \\
10. & $\epsilon_2$ & ms & 1.099 & 0.0003 & 1.204 & 0.0107 & 1.357 & 0.0285 \\  \hline
11. & $V_{t3}$ & mV & -65.38 & 0.0542 & -63.31 & 0.7557 & -62.20 & \textbf{2.037} \\
12. & $\delta V_3$ & mV & -17.65 & 0.0068 & -18.53 & 0.1935 & -20.79 & 0.6473 \\
13. & $\delta V_{\tau 3}$ & mV & 27.22 & 0.0136 & 26.82 & 0.2939 & 26.63 & \textbf{1.994} \\
14. & $t_3$ & ms & 0.7011 & 0.0035 & 0.8124 & 0.2339 & 1.077 & 0.6005 \\
15. & $\epsilon_3$ & ms & 12.90 & 0.0204 & 12.35 & 0.1620 & 11.98 & 0.2704 \\ \hline \hline
16. & $g_K$ & mS.cm$^{-2}$ & 6.901 & 0.0081 & 4.379 & 0.2335 & 3.046 & 0.0981 \\
17. & $E_K$ & mV & -99.99 & 0.0771 & -113.9 & \textbf{8.812} & -118.8 & \textbf{7.529} \\ \hline
18. & $V_{t4}$ & mV & -34.58 & 0.0069 & -35.76 & 0.2745 & -36.32 & 0.2053 \\
19. & $\delta V_4$ & mV & 22.17 & 0.0078 & 19.97 & 0.4515 & 14.85 & 0.5289 \\
20. & $\delta V_{\tau 4}$ & mV & 23.57 & 0.0102 & 22.20 & 0.3481 & 18.77 & 0.2407 \\
21. & $t_4$ & ms & 1.291 & 0.0081 & 1.038 & 0.2834 & 0.7844 & 0.1177 \\
22. & $\epsilon_4$ & ms & 4.315 & 0.0039 & 4.797 & 0.1185 & 7.165 & 0.1520 \\ \hline \hline
23. & $g_H$ & mS.cm$^{-2}$ & 0.1500 & 0.0190 & 0.1499 & \textbf{1.1751} & 0.1396 & \textbf{2.080} \\
24. & $E_H$ & mV & -42.991 & 0.1401 & -41.29 & \textbf{1.049} & -34.33 & \textbf{8.490} \\ \hline
25. & $V_{t5}$ & mV & -76.012 & 0.0227 & -75.89 & 0.6289 & -75.96 & 1.735 \\
26. & $\delta V_5$ & mV & -5.517 & 0.0531 & -6.923 & \textbf{1.006} & -9.162 & \textbf{3.374} \\
27. & $\delta V_{\tau 5}$ & mV & 20.278 & 0.2831 & 36.27 & \textbf{\textit{16.92}} & 55.35 & \textbf{\textit{30.57}} \\
28. & $t_5$ & ms & 6.445 & 0.7780 & 4.107 & \textbf{\textit{43.81}} & 4.742 & \textbf{\textit{38.13}} \\
29. & $\epsilon_5$ & ms & 54.85 & 0.7259 & 51.77 & \textbf{3.333} & 48.21 & \textbf{5.084} \\ \hline \hline
30. & $\bar{p}$ & $\mu$m.s$^{-1}$ & 0.1035 & 0.4509 & 0.2065 & \textbf{\textit{36.04}} & 0.0330 & \textbf{4.068} \\ \hline
31. & $V_{t6}$ & mV & -65.50 & 0.0191 & -65.42 & \textbf{2.098} & -63.48 & \textbf{\textit{16.48}} \\
32. & $\delta V_6$ & mV  & 12.40 & 0.0242 & 13.59 & \textbf{3.394} & 15.58 & \textbf{8.806} \\
33. & $\delta V_{\tau 6}$ & mV & 27.04 & 0.1222 & 19.25 & \textbf{9.183} & 13.49 & \textbf{\textit{10.42}} \\
34. & $t_6$ & ms & 0.7318 & 0.5071 & 2.126 & \textbf{\textit{35.76}} & 4.318 & \textbf{\textit{46.99}} \\
35. & $\epsilon_6$ & ms & 13.037 & 0.0509 & 27.97 & \textbf{\textit{32.26}} & 51.12 & \textbf{\textit{42.38}} \\ \hline
36. & $V_{t7}$ & mV & -86.00 & 0.2198 & -80.91 & \textbf{9.23} & -76.49 & \textbf{4.69} \\
37. & $\delta V_7$ & mV & -8.061 & 0.0316 & -4.811 & \textbf{3.24} & -1.995 & \textbf{2.56} \\
38. & $\delta V_{\tau 7}$ & mV & 16.72 & 0.1425 & 14.59 & \textbf{4.77} & 22.82 & \textbf{1.97} \\
39. & $t_7$ & ms & 28.11 & 0.27 & 32.18 & \textbf{8.497} & 34.79 & \textbf{\textit{17.59}} \\
40. & $\epsilon_7$ & ms & 288.1 & 0.5512 & 358.8 & \textbf{\textit{30.43}} & 287.2 & \textbf{\textit{36.73}} \\
\hline

\end{tabular}
\end{center}
\caption{\textbf{Dispersion of parameters inferred from erroneous models ErrM1 and ErrM2} \\
The $\mu_k$ and $\sigma_k$ are the mean values and standard deviations of parameters estimated from windows $\omega_1,\dots,\omega_{41}$ using the RVLM model (reference estimates).  The $\mu_{1k}$, $\sigma_{1k}$ and $\mu_{2k}$, $\sigma_{2k}$ are the means and standard deviations of parameters estimated with the erroneous models, ErrM1 and ErrM2 respectively. Figures in bold (bold+italics) single out deviations from mean greater than 1\% (10\% respectively).}
\label{tab:tab2}
\end{table*}

\subsubsection{Size of data set}

We have carried out assimilations using RPDA over larger windows of 20,001, 30,001, 40,001, and 49,999 points and found that the well-posed model converges in all of these cases.  The accuracy on parameter estimates is very similar to the accuracy reported in Table.\ref{tab:tab1}.  This suggests that larger windows do not produce further gains in accuracy in well-posed problems once they include a minimum of $\approx 5$ action potentials.

\subsection{Ill–posed problems}

We now study the dispersion of parameters inferred by four variants of the RVLM model which keep parametrization the same.  The $ErrM1$ variant had an erroneous gate exponent in the equation of the sodium current $J_{NaT}$ (Eq.\ref{eq:eqA2}) where the gate probability $x_2^3 x_3$ was replaced with $x_2^2 x_3$; the $ErrM2$ variant increased this error further by replacing $x_2^3 x_3$ with $x_2 x_3$; the $ErrM3$ variant was an over-specified RVLM model with an extra potassium A-channel; the $ErrM4$ variant combined model error with over-specification by adding the extra A-channel to the $ErrM1$ model.  Assimilating data with over-specified models was useful to determine the degree to which RPDA was able to filter out an unexpressed ion channel.  A hypothesis which we will be testing here is whether RPDA assigns a finite conductance to the A-channel to compensate for model error in the Na gate probability, or whether RPDA succeeds in disentangling the contributions of each ion channel \textit{in spite of} model error.

\begin{figure*}
\includegraphics[width=\linewidth]{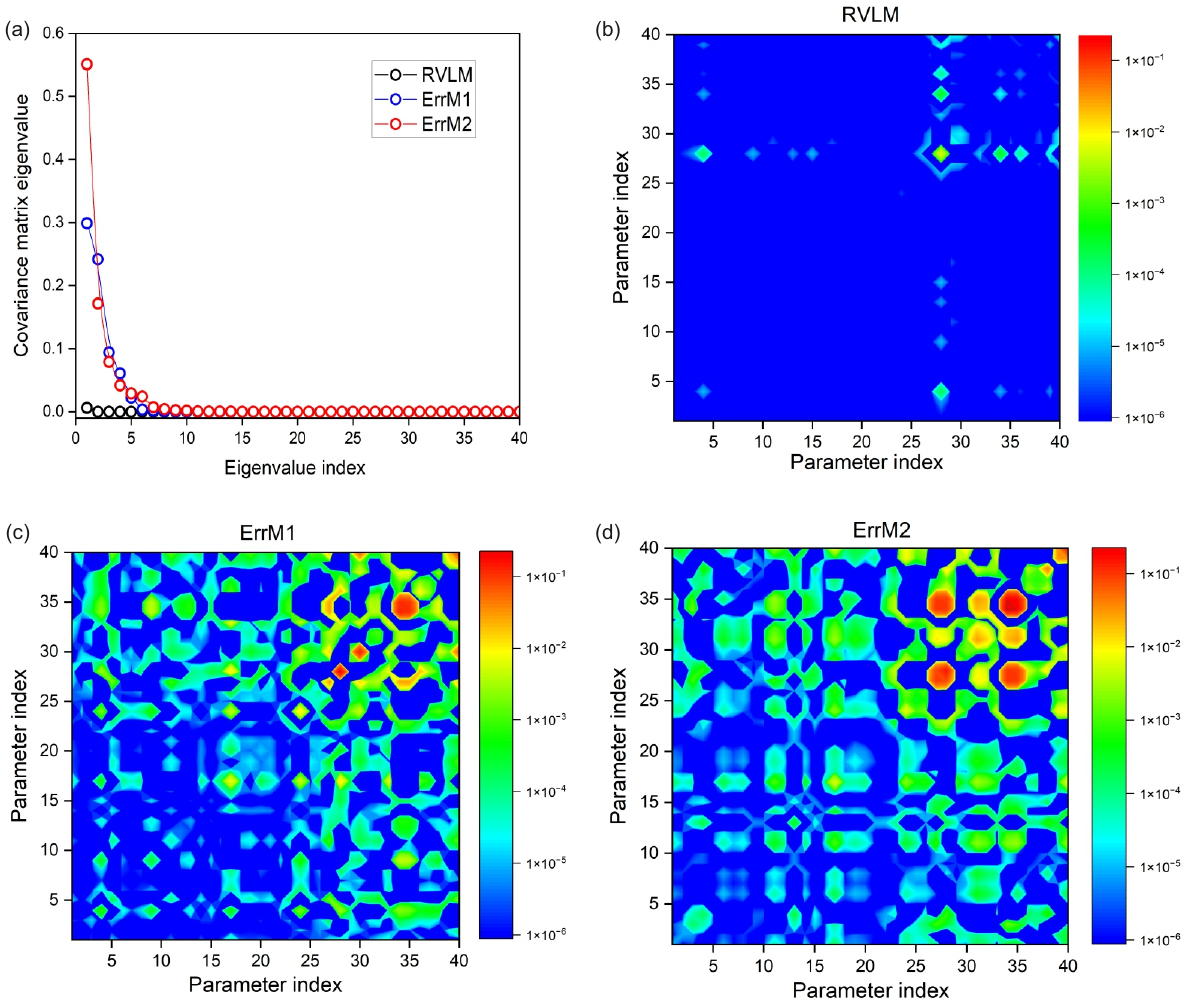}\\
\caption{\textbf{Covariance maps of parameters inferred with the RVLM, ErrM1 and ErrM2 models}
\\
(a) Eigenvalues of the covariance matrix of parameters estimated with erroneous models ErrM1 (blue trace) and ErrM2 (red trace).  The eigenvalues of the well-posed RVLM model (black trace) are shown for reference.  Covariance maps of the (b) RVLM, (c) ErrM1, and (d) ErrM2 parameters.}
\label{fig:fig3}
\end{figure*}

For each erroneous model, we estimated 41 sets of parameters from 41 assimilation windows, $\omega_1,\dots,\omega_{41}$ in Fig.\ref{fig:fig2}.  We then calculated the mean values and standard deviations of the $ErrM1$ and $ErrM2$ parameters which we compared to the true parameters values of the RVLM model (Table~\ref{tab:tab2}).  The covariance matrices of the RVLM, $ErrM1$ and $ErrM2$ parameters are plotted in Fig.\ref{fig:fig3}.  These show the emergence of parameter correlations as soon as model error is switched on in $ErrM1$ and $ErrM2$.  We then completed the $ErrM1$...$ErrM4$ models with the parameters estimated by RPDA and forward-integrated these models to predict the Na, K, HCN and A-type current waveforms (Fig.\ref{fig:fig4}).  These are compared to the true waveforms of the original RVLM model (dashed lines) to determine the error in the currents predicted by erroneous models.  The degree of confidence on predictions was determined by computing the min and max current waveforms (Fig.\ref{fig:fig4}) and the standard deviations on the integral charge under these curves (Table~\ref{tab:tab3}).  The Na, K and A-type current waveforms were reconstructed at the site of one and the same action potential (labelled * in Fig.\ref{fig:fig2}) which was chosen for being common to all 41 assimilation windows used to generate the statistical sample of parameters.

% The computation of ionic currents effectively integrates the parameter correlations caused by model error.  As a result, the prediction on ionic currents carry a higher degree of confidence than parameters estimates.  The predicted currents are also closer to the true values (Table~\ref{tab:tab3}).

\begin{table*}[t]
\begin{center}
\begin{tabular}{|c c|c c|c c|c c|c c|c c|c c|}
\cline{3-14}
\multicolumn{2}{c|}{ } & \multicolumn{2}{c|}{Well-posed} & \multicolumn{10}{c|}{Ill-posed} \\
\cline{3-14}
\multicolumn{2}{c|}{ } & \multicolumn{2}{c|}{RVLM} & \multicolumn{2}{c|}{ErrM1} & \multicolumn{2}{c|}{ErrM2} & \multicolumn{2}{c|}{ErrM3} & \multicolumn{2}{c|}{ErrM4} & \multicolumn{2}{c|}{Noisy} \\
\cline{3-14}
\multicolumn{2}{c|}{ } & \;\; $\bar{Q}$ \;\; & $ \;\; \frac{\Sigma}{\bar{Q}}$ \;\; & \;\; $\bar{Q}_1$ \;\; & \;\; $\frac{\Sigma_1}{\bar{Q}_1}$ \;\; & \;\; $\bar{Q}_2$ \;\; & \;\; $\frac{\Sigma_2}{\bar{Q}_2}$ \;\; & \;\; $\bar{Q}_3$ \;\; & \;\; $\frac{\Sigma_3}{\bar{Q}_3}$ \;\; & \;\; $\bar{Q}_4$ \;\; & \;\; $\frac{\Sigma_4}{\bar{Q}_4}$ \;\; & \;\; $\bar{Q}_N$ \;\; & \;\; $\frac{\Sigma_N}{\bar{Q}_N}$ \;\; \\
\multicolumn{2}{c|}{ } &  & \scriptsize \% & \scriptsize \% & \scriptsize \% & \scriptsize \% & \scriptsize \% & \scriptsize \% & \scriptsize \% & \scriptsize \% & \scriptsize \%  & \scriptsize \% & \scriptsize \% \\
\cline{3-14}
\noalign{\vskip 0.9mm} \hline
$Q_{Na}$ & nC.cm$^{-2}$ & 1126 & \textbf{1.06} & 1186 & 0.25 & 1224 & \textbf{4.50} & 1148 & \textbf{1.48} & 1166 & \textbf{1.24} & 1119 & 0.98 \\
$Q_{K}$ & nC.cm$^{-2}$ & 1098 & \textbf{1.54} & 1147 & 0.26 & 1174 & \textbf{3.32} & 1113 & 0.45 & 1131 & 0.97 & 1091 & \textbf{1.47} \\
$Q_{HCN}$ & nC.cm$^{-2}$ & 260 & \textbf{3.46} & 266 & \textbf{2.69} & 291 & \textbf{4.85} & 263 & \textbf{2.28} & 263 & \textbf{1.90} & 258 & \textbf{3.49} \\
%$Q_{A}$ & nC.cm$^{-2}$ & 0 & 0 & - & - & - & - & (-82) & 123 & (-120) & 124 & - & -  \\
\hline
\end{tabular}
\end{center}
\caption{\textbf{Na, K and HCN ion discharge per action potential} \\ $Q_{Na}$, $Q_{K}$, and $Q_{HCN}$ are the ionic volumes discharged during action potential (*).  $\bar{Q}_1$ and $\Sigma_1$ are the mean values and standard deviations derived from a statistical sample of 41 ion current waveforms.  These waveforms are predicted by 41 $ErrM1$ models completed with the 41 parameter sets from assimilation windows $\omega_1,\dots,\omega_{41}$.  The same was done for RVLM, $ErrM2$, $ErrM3$ and $ErrrM4$. $\bar{Q_N}$ and $\Sigma_N$ are mean charge and standard deviations obtained by assimilating noisy data with the RVLM model.}
\label{tab:tab3}
\end{table*}

\subsubsection{Erroneous model variants: $ErrM1$ and $ErrM2$}

Detuning the gate exponent from $x_2^3 x_3$ (RVLM) to $x_2^2 x_3$ ($ErrM1$) and $x_2 x_3$ ($ErrM2$) introduces correlations between parameters.  Their standard deviations jump by an order of magnitude from RVLM to $ErrM1$ (Table~\ref{tab:tab2}) and by a smaller increment from $ErrM1$ to $ErrM2$.  Gate recovery times previously identified as the least well constrained parameters such as $t_5$,$\epsilon_5$ or $t_6$, $\epsilon_6$ are the most sensitive to model error.  The ($t_6$,$\epsilon_6$) pair is an extreme example where standard deviations jump from (0.5\%,0.05\%) (RVLM), to (35\%,32\%) ($ErrM1$), and (47\%,42\%) {$ErrM2$}.

Parameter correlations are also evident in the finite off-diagonal components of the $ErrM1$ and $ErrM2$ covariance matrices (Fig.\ref{fig:fig3}) compared to the RVLM covariance matrix which has none - except for $t_5$ (parameter 28).  The eigenvalues of the RVLM covariance matrix (Fig.\ref{fig:fig3}a) are vanishingly small.  In contrast, the $ErrM1$ and $ErrM2$ covariances have 6 non-zero eigenvalues.  These indicate the lengths of the principal semi-axes along which parameter compensation occurs.  The largest correlations are within the parameters of the CaT and HCN currents ($k>25$) and they increase with increasing model error (Fig.\ref{fig:fig3}b-d).  An important point to note is the clustering of off-diagonal terms in blocks linking the parameters of individual ionic currents.  This raises the prospect that parameter correlations may compensate each other in the calculation of ionic currents to make more accurate predictions than by relying on parameters alone.

\begin{figure*}
\includegraphics[width=\linewidth]{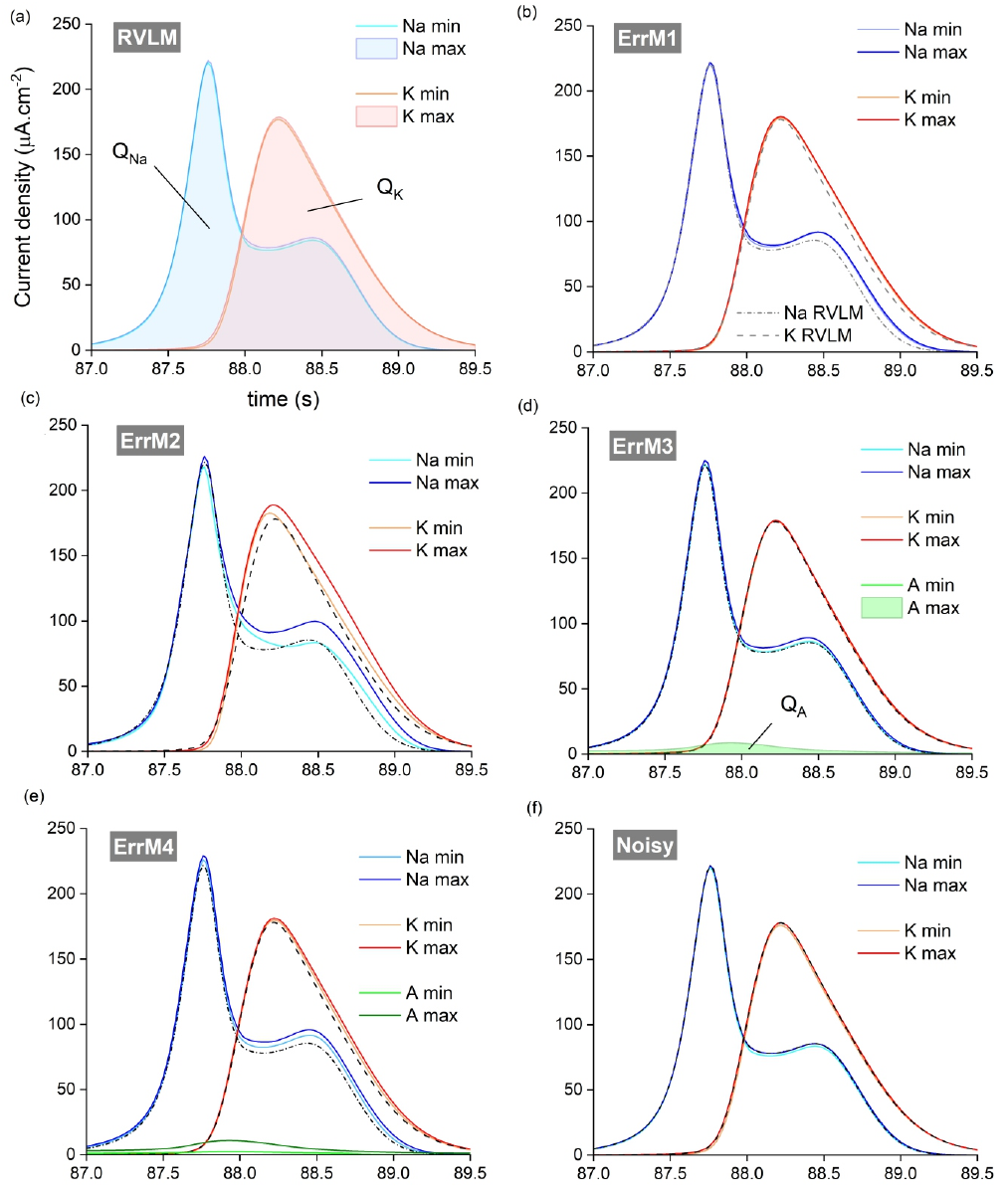}\\
\caption{\textbf{A comparison of ionic currents predicted by erroneous and exact models}
\\
(a) Current waveforms predicted by the $RVLM$ model.  The min and max in the Na and K traces give the range of variation of currents reconstructed from the 41 assimilation windows.  For reference, the true Na and K current waveforms predicted by the original $RVLM$ model are shown as the black dashed lines in (b-f).  Current waveforms predicted by erroneous models: (b) $ErrM1$, (c) $ErrM2$; an over-specified $RVLM$ model: (d) $ErrM3$; an over-specified erroneous model: (e) $ErrM4$; and from noisy data, (f) $Noisy$.}
\label{fig:fig4}
\end{figure*}

Fig.\ref{fig:fig4} shows the range of current waveforms reconstructed from our 41 sets of 40 parameters.  For clarity we only plot the min and max waveforms of this range.  The min and max current waveforms predicted by the RVLM model are virtually identical to the original waveforms (Fig.\ref{fig:fig4}a).  Quite remarkably, the $ErrM1$ model still predicts nearly identical min and max waveforms (Fig.\ref{fig:fig4}b) despite the order of magnitude larger uncertainty in the underlying parameters (Table~\ref{tab:tab2}).  Only in the $ErrM2$ model does the dispersion in predicted currents become noticeable (Fig.\ref{fig:fig4}c).  Note also the excellent agreement between the currents predicted by $ErrM1$ and $ErrM2$ and the true RVLM current waveform (dashed lines).  These results demonstrate that approximate models can still predict the true currents with a high degree of confidence despite large uncertainty in underlying parameters.

We then calculated the total Na and K charge transferred under the predicted current waveforms.  The mean values $\bar{Q}$ and standard deviations $\Sigma$ are given in Table~\ref{tab:tab3}.  The standard deviations $\Sigma_1/\bar{Q}_1$ ($ErrM1$) and $\Sigma_2/\bar{Q}_2$ ($ErrM2$) remain under 5\% for all currents.  This is considerably lower than the uncertainty on underlying parameters which can be as high as 50\% (Table~\ref{tab:tab3}).  If we examine how close the average estimates are to true values we find that the $\bar{Q}_1$ is within 5\% (Na), 4\% (K) of the original model prediction and $\bar{Q}_2$ is within 9\% (Na), 7\% (K).  The ionic charges predicted by erroneous models $ErrM1$ and $ErrM2$ are therefore an excellent predictor of the true ionic charges.  In our example, decreasing the Na gate exponent effectively decreases the slope of the activation curve.  RPDA compensates for the subsequent widening of the activation curve by reducing the width parameter $\delta V_2$ from 10 (RVLM) to 8.1 ($ErrM1$) and 5.2 ($ErrM2$) in Table~\ref{tab:tab2}.  The large deviation (50\%) of such parameter from true value is a second reason why ionic currents constitute more stable predictors of true values when the model is unknown.

\subsubsection{Over-specified models: $ErrM3$ and $ErrM4$}

It is sometimes argued that model over-specification causes multi-valued solutions.  We investigate this hypothesis by constructing two over-specified models, adding a super-numerary A-type current to the RVLM model ($ErrM3$), and to the $ErrM1$ model ($ErrM4$).  The A-type current density is $J_{A} = g_{A}\, x_{14} \, x_{15} \, (x_1-E_{K})$, where $x_{14}$ and $x_{15}$ are the activation and inactivation variables respectively, and $E_{K}$ is the potassium reversal potential.

We find that the $ErrM3$ problem has a single-valued solution that yields the true RVLM parameters.  In the process, RPDA correctly filters out the A-type current by assigning a negligible value to $g_A$.  The standard deviations of the $ErrM3$ parameters are also small ($<0.01\%$), and comparable to those of the RVLM model (Table~\ref{tab:tab2}).  The narrow dispersion of parameter estimates explains the narrow dispersion of the predicted current waveforms and their similarity to the true current waveforms (dashed lines, Fig.\ref{fig:fig4}d).  The mean ion volume discharged per action potential, $\bar{Q}_3$, is within 1.5\% (Na) and 3\% (K) of true values hence is also an excellent predictor of true ion discharge.  The standard deviations, $\Sigma_3$, are similar to those of the RVLM model (Table~\ref{tab:tab3}).  These results show that large models still obtain the correct parameters and currents despite over-specification.

Now turning to the $ErrM4$ model, one hypothesis is that the A-channel parameters might compensate for the erroneous Na gate exponent.  Examination of parameter distributions show nothing of the sort occurs.  RPDA assigns a vanishing conductance, $g_A$, to the supernumerary `A-channel'.  This can be seen in Fig.\ref{fig:fig4}e where the A-current is zero within a standard deviation.  The predicted Na and K current waveforms remains in excellent agreement with true waveforms (dashed lines) and carry a high degree of confidence \textit{in spite of model error}.  Turning to the mean charge transferred per action potential, $\bar{Q}_4$, we find $ErrM4$ predicts the true ionic discharge within 3.5\% (Na) and 3\% (K) of true values.  The volume of A-type charge transfer, $\bar{Q}_4$=82nC.cm$^{-2}$ remains tiny compared to 1166nC.cm$^{-2}$ for Na and 1131nC.cm$^{-2}$ for K indicating the A-channel does not compensate for gate exponent error.   The standard deviations on discharged ions, $\Sigma_4$, are small and comparable to those of the RVLM model (Table~\ref{tab:tab3}).

These remarkable results indicate that RPDA successfully disentangles the contributions of all ion current types \textit{even with approximate models that may include redundant ion channels}.

\subsubsection{Data error: assimilating noisy RVLM data}

For completeness we examine the case where the $V_{mem}$ time series in Fig.\ref{fig:fig2}a is corrupted by additive Gaussian noise.  The noise r.m.s. amplitude, $0.1$ mV, is comparable to the noise level in patch-clamp recordings.  Fig.\ref{fig:fig4}f shows the Na and K currents predicted by the RVLM model in this case.  The current waveforms have a very narrow dispersion and are virtually identical to the true waveforms (dashed lines).  Hence the impact of data error on predicted quantities will often be small compared to model error.

\subsubsection{Unknown model: hippocampal CA1 neuron}

\begin{table}[t]
\begin{center}
\begin{tabular}{|c c|c|c|}
\cline{3-4}
\multicolumn{2}{c|}{ } & \multicolumn{2}{c|}{CA1 neuron} \\
\cline{3-4}
\multicolumn{2}{c|}{ } & \;\; $\bar{Q}$ \;\; & $ \;\; \frac{\Sigma}{\bar{Q}}$ \;\; \\
\multicolumn{2}{c|}{ } &  & \scriptsize \% \\
\cline{3-4}
\noalign{\vskip 0.9mm} \hline
$Q_{Na}$ & nC.cm$^{-2}$ & 73.43 & 14.4 \\
$Q_{K}$ & nC.cm$^{-2}$ & 466.2 & 14.6 \\
\hline

\end{tabular}
\end{center}
\caption{\textbf{Na and K ion discharge per action potential of CA1 neuron} \\  Mean ion volumes discharged, $Q$, and standard deviation, $\Sigma$, at action potential (*) in Fig.\ref{fig:fig5}.}
\label{tab:tab4}
\end{table}

\begin{figure*}
\includegraphics[width=\linewidth]{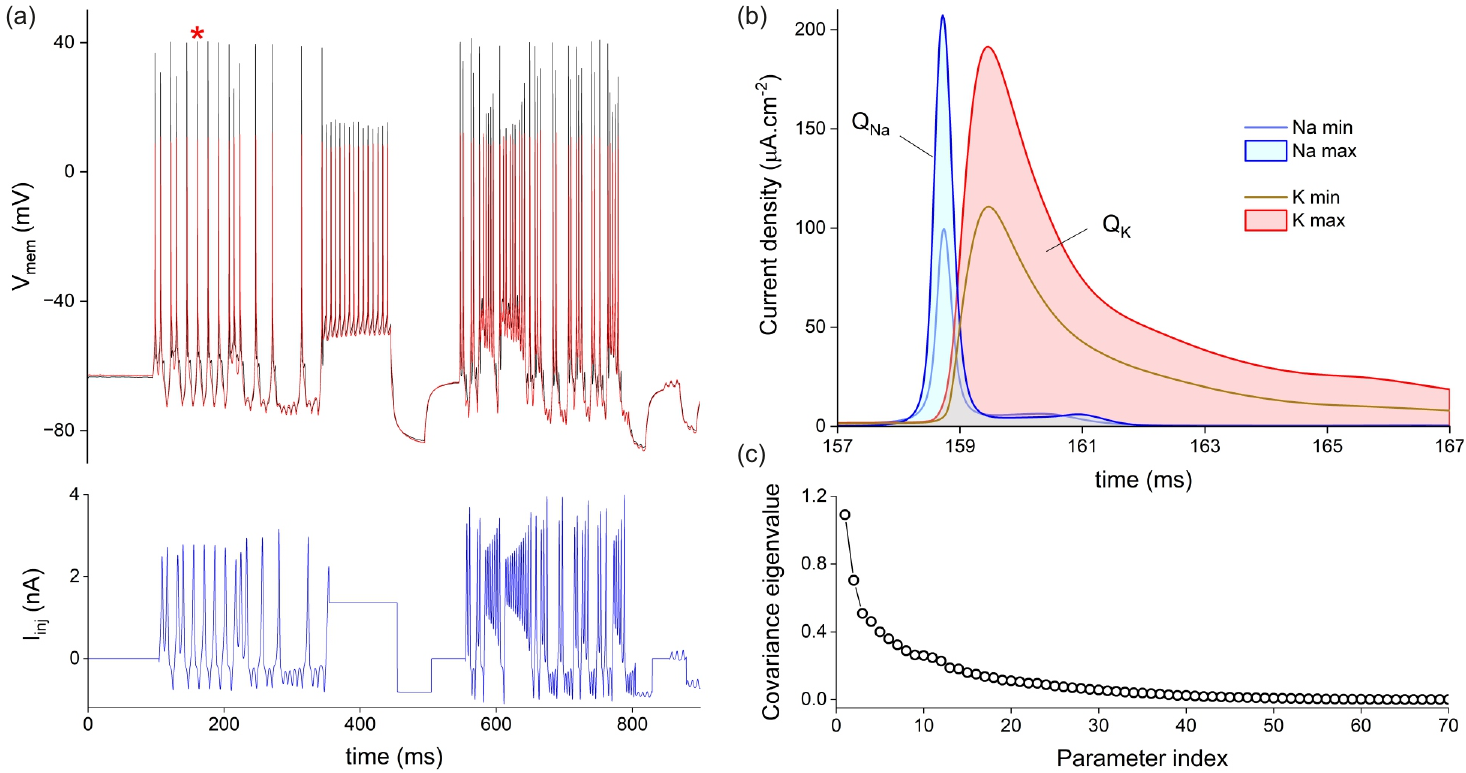}\\
\caption{\textbf{Uncertainty on currents and parameters estimated from a hippocampal CA1 neuron}
\\
(a) Membrane voltage oscillations (black trace) driven by injecting a calibrated current waveform (blue trace) into a hippocampal neuron (Wistar rat, CA1 neuron).  Voltage oscillations predicted by the hippocampal neuron model completed with one set of parameters.  The (*) labels the action potential whose ionic current waveforms we predict in (b).  (b) Mini and maxi current waveforms (Na, K) predicted by 71 CA1 models completed with the parameters from 71 assimilation windows.  (c) The eigenvalues of the $70 \times 70$ covariance matrix of estimated parameters.}
\label{fig:fig5}
\end{figure*}

We complete our study of model error by investigating the distribution of parameters and ionic currents estimated using a guessed neuron model.  Model error is now unknown which means that the mean parameter and current estimates can no longer be compared to the biological values which we are seeking.  We used the model to assimilate the current-clamp recordings of a hippocampal CA1 neuron (Fig.\ref{fig:fig5}a).  We refer the reader to Abu-Hassan et al.~\cite{AbuHassan2019} for details on the experimental protocol.  The guessed CA1 model had 9 ion channels and 70 parameters (see Appendix B).  It includes the NaT, NaP, K, A-type, Ca, BK, SK, HCN and leak ion channels believed to be present in the CA1 soma~\cite{Warman1994,Traub1991,Karst1993}.  We obtained 71 sets of 70 parameters from sliding time windows.  Each window was 200ms long.  Consecutive windows were offset by 2ms, same as in Fig.\ref{fig:fig2}.  The CA1 models completed with any of the 71 sets of parameters successfully predicted the observed membrane voltage (red trace, Fig.\ref{fig:fig5}a).  The good predictive power of the completed models therefore suggests that the error in the model equations is mild.

From these 71 sets of parameters, we reconstructed the ionic current waveforms at the site of the action potential labelled (*) in Fig.\ref{fig:fig5}a.  The predicted sodium currents (NaP, NaT) and potassium currents (K, A, SK, BK) are combined into Na and K waveforms in Fig.\ref{fig:fig5}b.  While the waveform shapes are similar to those of our earlier models (Fig.\ref{fig:fig4}) the predictions cover a wider range and the potassium current has a slower rate of decay (Fig.\ref{fig:fig5}b).  The standard deviations on the ionic charge transferred per action potential are 14.4\% (Na) and 14.6\% (K) (Table~\ref{tab:tab4}).  These are 7 times larger than in the $ErrM2$ model suggesting that the discrepancy between the guessed CA1 model and the actual CA1 neuron is approximately 7 times greater than the discrepancy between the $ErrM2$ model and the RVLM model.  Another measure of model error are the eigenvalues of the parameter covariance matrix (Fig.\ref{fig:fig5}c).  Although the CA1, $ErrM1$ and $ErrM2$ eigenvalues fall in the same range (Figs.\ref{fig:fig3}b,c;\ref{fig:fig5}c) the CA1 distribution has a fatter tail of eigenvalues which suggests the CA1 model has more parameters with small functional overlap.

\section{Discussion}

In well-posed problems, the RPDA method achieves a remarkable 100\% convergence rate from 28 different initial conditions (Table~\ref{tab:tab1}) and with 41 current waveforms (Table~\ref{tab:tab2}).  Improvement over the 64\% convergence rate reported by Taylor et al.~\cite{Taylor2020} is achieved by re-injecting data in piecewise intervals and by recursively restarting parameter search from a larger piecewise interval if convergence fails.  The well-posed case was used to validate the fulfilment of the identifiability and observability criteria when the true parameter solution was recovered starting from different initial conditions and current waveforms.  With the caveat that a solution to such NP-hard inference problem is not known to exist in general, our empirical simulations suggest that in the special case of neuron-based conductance models the RPDA method obtains a solution for systems of up to at least 14 differential equations which describe most neuron types.  The RVLM model is exemplar in this regard because any ionic current which may be added will have a similar mathematical form including sigmoidal activation and first order gate dynamics.

By intentionally introducing model error in well-posed data assimilation ($ErrM1$, $ErrM2$), multi-valued solutions were found to occur due to correlations between parameters (Fig.\ref{fig:fig3}).  Averaging the parameters estimated from sliding assimilation windows gave mean parameter estimates within 15\% of true values except for gate activation times which deviated by a factor of two or more and had large uncertainty (up to $46\%$).  In order to achieve a greater consistency in predictions, we calculated the ionic currents which integrate parameter correlations (Fig.\ref{fig:fig4}).  The mean ionic charge transferred per action potential deviated by less than 5\% ($ErrM1$) and 8\% ($ErrM2$) from true values and the coefficients of variation were very low at 0.26\% ($ErrM1$) and 4.5\% ($ErrM2$).  It is therefore anticipated that ionic currents could be predicted with sufficient accuracy to resolve changes in individual ionic currents induced by inhibitory drugs or ion channel dysfunction.

Overspecifying conductance models ($ErrM3$, $ErrM4$) is no impediment to recovering the true parameters.  The mean current waveforms estimated by $ErrM3$ are identical to those of the RVLM model and their coefficient of variation are identical ($<1.5\%$).  The $ErrM3$ waveforms remain close to the true waveforms (dashed lines, Fig.\ref{fig:fig4}) and their coefficient of variation is also small ($<1.2\%$).  The charge transferred under the $ErrM4$ waveforms was within 4\% of true values and the uncertainty on predictions was less than 1.3\% despite model error and overspecification.  This demonstrates the ability of the RPDA method to disentangle the correct contribution of each ion channel to the membrane voltage without the overspecified current compensating for model error.  The implication is that a universal conductance model could infer accurate information on ion channels without requiring any prior assumption on which ion channel might be expressed.  This also suggests that reductionist models that minimize the number of parameters may not be the best route to improve accuracy on parameter estimates.  Instead, focus should be on mitigating model error.

We found that standard deviations on parameters (Tables~\ref{tab:tab2}) and current estimates (Tables~\ref{tab:tab3},~\ref{tab:tab4}) increase with increasing model error.  Therefore standard deviations provide a metric to quantify how close a guessed model is to the unknown biological model.

A possible strategy for correcting model error has been proposed by Ye et al.~\cite{Ye2015,Abarbanel2022}.  This consists in adding a model error term to the cost function and balancing data error and model error in such a way as to eliminate the bias of model error on the parameter solution.  When model error is properly weighted, parameters are assigned true values without having to compensate for model error.  The challenge is in the determination of the covariance matrix representing this weight~\cite{Ye2015}.   We have attempted to include a model error term in Eq.\ref{eq:eq3} and weighted it with a single hyperparameter.  However this approach has proved unpractical due to the difficulty of finding a finite hyper-parameter value that resolves all constraint violations.  The optimal value of such hyperparameter was found to drift to zero or infinity during parameter search.  Further progress will require evaluating the covariance matrix weighting model error.

\section{Conclusions}

We have introduced Recursive Piecewise Data Assimilation as an novel method for optimizing neuron-based conductance models. RPDA improves over earlier parameter estimation methods by biasing the parameter search with data and by iteratively improving the solution as the bias is gradually released.  When the model is known, RPDA achieves a 100\% convergence rate towards the true solution for a wide range of initial conditions and training data sets.  When the model is unknown, model error introduces correlations between parameters estimates.  The largest correlations occur between parameters that define each ionic current.  By reconstructing the ionic currents, parameter correlations cancel out and the current waveforms are found to approximate very well the true current waveforms.  The ionic charge transferred under these curves is also predicted with a high degree of confidence of 95.5\% when model error affects a single ion channel and 85\% when the true biological model is guessed (CA1 neuron).  The increasing uncertainty of predictions with increasing model error suggests that the covariance matrix of parameters is a good metric of model error.  We also found that model over-specification is not responsible for parameter sloppiness and that RPDA correctly disentangles all ion channel contributions to the membrane voltage data even when the model has a wrong gate exponent.  Our work shows that combining variational inference with statistical analysis offers good prospects for extracting sensible biological information from data.

\section{Appendix A: RVLM neuron model}

Our model of the Rostro-Ventro-Lateral Medulla (RVLM) considers the soma as a single compartment~\cite{Moraes2013,Taylor2020} with 5 types of ionic currents: transient sodium (NaT), delayed-rectifier potassium (K), low threshold calcium (CaT), hyperpolarisation-activated cation (HCN), and a leakage current (L).  The model has 7 state variables ($L=7$) and 40 adjustable parameters ($K=40$).  The neuron membrane voltage $x_1$ varies as:

\begin{equation}
C \frac{dx_1}{dt} = -J_{NaT} -J_{K} -J_{CaT} -J_{HCN} -J_L + \frac{I_{inj}(t)}{A},
\label{eq:eqA1}
\end{equation}

\noindent when driven by injected current $I_{inj}$.  $A$ is the effective area of the soma.  The ionic current densities are:

\begin{equation}
\begin{aligned}
&J_{NaT} = g_{NaT} x_2^3 x_3 (x_1-E_{Na}), \\
&J_{K} = g_{K} x_4^4 (x_1-E_{K}), \\
&J_{HCN} = g_H \, x_5 \, (x_1-E_H), \\
&J_{CaT} = 4\bar{p} x_6^2 x_7 \frac{x_1 F^2}{RT}\frac{[Ca^{2+}]_{i}-[Ca^{2+}]_{o}e^{-2F x_1 /RT}}{1-e^{-2F x_1 /RT}}, \\
&J_{L} = g_{L} (x_1-E_{L}).
\end{aligned}
\label{eq:eqA2}
\end{equation}

\noindent Nominal values of reversal potentials are: $E_{Na}$ = +41mV, $E_{K}$ = -100mV, $E_{H}$ = -43mV, $E_{L}$ = -65mV; ionic conductances: $g_{NaT}$ = 69.0 mS.cm$^{-2}$, $g_{K}$ = 6.90 mS.cm$^{-2}$, $g_{H}$ = 0.15 mS.cm$^{-2}$, $g_{L}$ = 0.465 mS.cm$^{-2}$; maximum calcium permittivity: $\bar{p}$=0.103$\mu$m.s$^{-1}$; intracellular and extracellular calcium concentrations: $[Ca^{2+}]_i=2.4\times10^{-10}$mol.cm$^{-3}$ and $[Ca^{2+}]_o=2.0\times10^{-10}$mol.cm$^{-6}$.  $F=9.65\times10^4$C.mol$^{-1}$ is Faraday's constant, $R=8.324$J.K$^{-1}$.mol$^{-1}$, the ideal gas constant, and $T=298$K.

Ionic gates follow a first order dynamics:

\begin{equation}
\frac{dx_l}{dt} = \frac{x_{l,\infty}(x_1) - x_l}{\tau_l(x_1)}, \;\;l=2 \dots 7
\label{eq:eqA3}
\end{equation}

\noindent where $x_l \in \left\{V,m,h,n,z,q,r\right\}$ and:

\begin{equation}
\begin{aligned}
&x_{l,\infty}(x_1) = \frac{1}{2}\left( 1 + \tanh\frac{x_1 - V_{t,l}}{\delta V_l} \right), \\
&\tau_{l}(V) = t_l + \epsilon_l \left( 1 - \tanh^2 \frac{x_1-V_{t,l}}{\delta V_{\tau,l}} \right).
\end{aligned}
\label{eq:eqA4}
\end{equation}

\noindent The nominal RVLM parameters are listed in the ``True" value column of Table~\ref{tab:tab1}.  Eqs.\ref{eq:eqA1}-\ref{eq:eqA4} were configured with these parameters to generate the membrane voltage data $V_{mem}$ in  Fig.\ref{fig:fig2}.  The injected current waveform $I_{inj}(t)$ had 50,000 points with a 0.02 ms time step.  The RVLM model was forward-integrated with the \emph{odeint} function of the \emph{scipy} package in Python 3.  The resulting $V_{mem}$ data were then assimilated by the RVLM system and its $Errm1$-$ErrM4$ variants.

\section{Appendix B: Hippocampal neuron model}

This hippocampal neuron model includes the ion channels in the soma of CA1 neurons~\cite{Golomb2006,Yue2005,Karst1993,Traub1991,Warman1994}: persistent sodium current (NaP), a A-type potassium current (A) and calcium activated potassium currents (SK and BK) in addition to the NaT, K, Calcium, HCN and leak currents of the previous RVLM model.  The model now has 9 ionic currents, 14 state variables ($L=14$) and 69 adjustable parameters ($K=69$).  The membrane voltage dynamics is given by:

\begin{eqnarray}
C \frac{dx_1}{dt}& = & -J_{NaT} -J_{NaP} -J_{K} -J_{A} -J_{Ca} \\
&  & -J_{BK} -J_{SK} -J_{HCN} -J_L + \frac{I_{inj}(t)}{A}.
\label{eq:eqB1}
\end{eqnarray}

\noindent The ionic current densities are:

\begin{equation}
\begin{aligned}
&J_{NaT} = g_{NaT} x_2 x_3^3 (x_1-E_{Na}), \\
&J_{NaP} = g_{NaP} x_4 (x_1-E_{Na}), \\
&J_{K} = g_K x_5^4 (x_1-E_{K}), \\
&J_{A} = g_A x_6 x_7 (x_1-E_{K}), \\
&J_{Ca} = g_{Ca} x_8^2 x_9 (x_1-E_{Ca}), \\
&J_{BK} = g_{BK} x_{10}^2 x_{11} (x_1-E_{K}), \\
&J_{SK} = g_{SK} x_{12} (x_1-E_{K}), \\
&J_{HCN} = g_{H} x_{13} (x_1-E_{HCN}), \\
&J_{L} = g_{L} (x_1-E_{L}).
\end{aligned}
\label{eq:eqB2}
\end{equation}

The BK current depends on both the membrane voltage $x_1$ and the internal calcium concentration $[Ca^{2+}]_{i}\equiv x_{14}$ whereas the SK current depends on $[Ca^{2+}]_{i}$ only.  The BK current has fast activation which we take to be instantaneous in line with Warman et al.~\cite{Warman1994}:

\scriptsize
\begin{equation}
x_{10,\infty}= 0.5\left[1+\tanh\left\{x_1-V_{10}+130\frac{1+\tanh\frac{x_{14}}{0.2}-250}{\delta V_{10}}\right\}\right],
\label{eq:eqB3}
\end{equation}
\normalsize

\noindent where the calcium equilibrium across the membrane is given by the following rate equation:

\begin{equation}
\frac{dx_{14}}{dt} = \frac{x_{14,\infty}-x_{14}}{\tau_{11}}-\frac{J_{Ca}}{4w},
\label{eq:eqB4}
\end{equation}

\noindent driven by the calcium ion current $J_{Ca}$.  $w$ is the thickness of the surface area across which Ca$^{2+}$ fluxes are calculated ($w=1\mu$m).  We modelled the dynamics of calcium gate variables $x_8$ and $x_9$ with Eqs.\ref{eq:eqA3} and \ref{eq:eqA4} with the appropriate calcium gate parameters.

\noindent The inactivation dynamics of the BK current has a long time constant $\tau_{11}$.  It also follows rate equation:

\begin{equation}
\frac{dx_{11}}{dt} = \frac{x_{11,\infty}-x_{11}}{\tau_{11}},
\label{eq:eqB5}
\end{equation}

\noindent where

\scriptsize
\begin{equation}
x_{11,\infty} = 0.5\left[1+\tanh\left\{x_1-V_{11}+130\frac{1+\tanh\frac{x_{14}}{0.2}-250}{\delta V_{11}}\right\}\right]
\label{eq:eqB6}
\end{equation}
\normalsize

The activation dynamics of the SK current is fast ($x_{12} \equiv x_{12,\infty}$), and similar to the BK channel, the activation of the SK channel is described by:

\scriptsize
\begin{equation}
x_{12,\infty} = 0.5\left[1+\tanh\frac{x_{14}-x_{t,12}}{\delta x_{\tau,12}}\right].
\label{eq:eqB7}
\end{equation}
\normalsize

Gate variables $x_2$ (NaP channel) and $x_6$, $x_7$ (A channel) follow the first order response in Eqs.\ref{eq:eqA3} and \ref{eq:eqA4} with the appropriate NaP and A channel parameters.

\section{Acknowledgements}

This work was supported by the European Union's Horizon 2020 Future Emerging Technologies Programme (Grant No. 732170).

\section{Ethical statement}

Experiments on rodents were performed under Schedule 1 in accordance with the United Kingdom Scientific procedures act of 1986.

\bibliography{biblioMSS.bib}

\end{document}